\documentclass[11pt]{article}
\usepackage{fullpage,amsthm,amsmath,times,color}
\usepackage{graphicx,epstopdf,subfigure,comment}
 \usepackage{afterpage}

\begin{document}
\newcommand{\ie}{\emph{i.e.}}
\newcommand{\eg}{\emph{e.g.}}
\newcommand{\ER}{Erd\H{o}s-R\'{e}nyi}
\global\long\def\Rv{\mathcal{R}_v}
\newcommand{\JN}[1]{{\color{magenta}  #1}}
\newcommand{\JNdel}[1]{{\color{green}  #1}}
\newcommand{\RD}[1]{{\color{red}  #1}}
\newcommand{\RDchanged}[1]{{\color{blue}  #1}}

\long\def\symbolfootnote[#1]#2{\begingroup%
\def\thefootnote{\fnsymbol{footnote}}\footnote[#1]{#2}\endgroup}

\begin{center}
{\Large {\bf  Explosive Percolation: Novel critical and supercritical phenomena} }\\[3mm]
\ \ \ \ \ \ \ \ {\it Raissa M. D'Souza \ \ \ \ \ \ \ \ \ \ \ \ \ \ \ \ \ \ \ \ \ \ \ \ \ \ \ \ \ \ \ \ \ \ \ \ \ \ \ \ \ \ \ \ Jan Nagler}\\
{\it University of California, Davis} \ \ \ \ \ \ \ \ \ \ \ \ \ \ \ \ \ \ \ \ \ \ \ \ \ \ \ \ \ \ \ \ {\it  ETH Zurich}\\
\ \ \ \ \ {\it  raissa@cse.ucdavis.edu} \ \ \ \ \ \ \ \ \ \ \ \ \ \ \ \ \ \ \ \ \ \ \ \ \ \ \ \ \ \ \ \ \ \ \ \ \ \ \ {\it jnagler@ethz.ch} 


\end{center}
\hrule
\vspace{0.2in}

Explosive Percolation describes the abrupt onset of large-scale connectivity that results from  a simple random process designed 
to delay the onset of the transition on an underlying random network or lattice. Explosive percolation transitions exhibit an array of novel universality classes and supercritical behaviors including a stochastic sequence of discontinuous transitions, multiple giant components, and lack of self-averaging. Many mechanisms that give rise to explosive percolation have been discovered, including overtaking, correlated percolation, and evolution on hierarchical lattices. Many connections to real-world systems, ranging from social networks to nanotubes, have been identified and explosive percolation is an emerging paradigm for modeling these systems as well as the consequences of small interventions intended to delay phase transitions. This review aims to synthesize existing results on explosive percolation and to identify fruitful directions for future research.

\section*{Introduction}

Percolation, the emergence of large-scale connectivity, 
is a theoretical underpinning across a range of fields~\cite{StaufferPercBook,SahimiBook}. 
The extent of connectivity that emerges suddenly at a critical point, $t_c$, has a profound impact on the macroscopic behaviors of a system. At times, ensuring large-scale connectivity is essential. For instance, a transportation network (like the world-wide airline network) or a communication system (like the internet) is only useful if a large fraction of the nodes can reach one another. Yet, in other contexts, large-scale connectivity is a liability.  For instance, a virus spreading on a social or computer network above $t_c$ can reach an extensive fraction of nodes causing an epidemic outbreak; below $t_c$, each outbreak would be contained in a small, isolated cluster. There is thus great interest in controlling the location of the percolation transition to either enhance or delay its onset and, more generally, in understanding the consequences of control interventions. Although the topic has been of interest for many years, it was only recently established that small delay interventions can have drastic, unanticipated and exciting consequences. 
Here we review Explosive Percolation (EP), 
the phenomenon that often results from repeated, small interventions designed to delay the percolation phase transition. The onset can indeed be significantly delayed, but once the percolation transition is inevitably reached, large-scale connectivity emerges dramatically.  

\section*{Random graph percolation}

\begin{figure}[tb]
  \begin{center}
   \includegraphics[width=.9\textwidth]{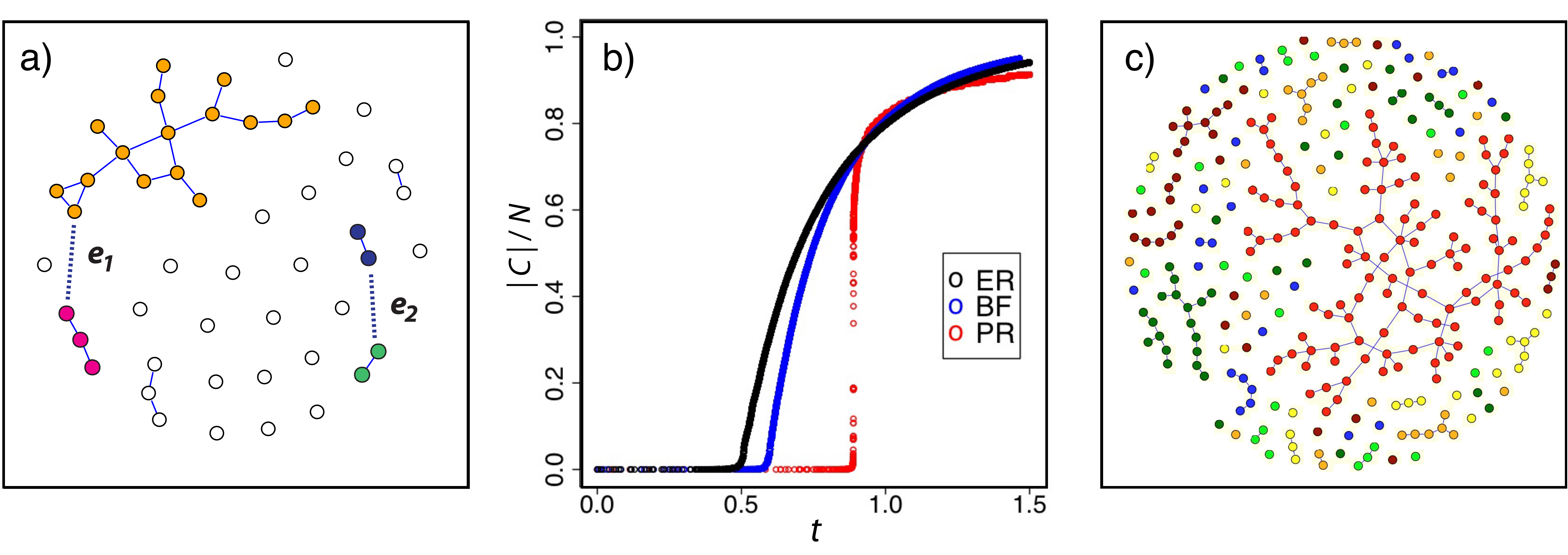}
  \end{center}
  \vspace{-0.2in}
  \caption{
  {\bf Schematic of Explosive Percolation.}
  (a)  At each time step of the Product Rule (PR) process, two edges, $e_1$ and $e_2$, compete for addition.
  Here the product of the components merged by $e_1$ is 
  $3\times 16=48$ and by $e_2$ is $2\times 2=4$,
  so $e_2$ is accepted in and $e_1$ rejected.
   (b) Typical evolution of an \ER~(ER), Bohman Frieze (BF), and PR process on a system of size $N=10^6$.
    Plotted is the fractional size of the largest component, $|C|/N$, as a function of edge density $t$.
  (c) A sample ER network in the supercritical regime with the nodes in each distinct component rendered in the same color. 
  The largest component, $C$, is indicated in red.
  }
\label{figBasics}
\end{figure}

{
Traditional percolation on a random graph, the \ER~(ER) model, considers a collection of $N$ isolated nodes, with each possible edge between two distinct nodes added to the graph with probability $p$~\cite{ER1959,ER,gilbert1959}.
This is a static formulation, parameterized by $p$, with no dependence on the history of how edges have been added to the graph. In contrast, we consider a mathematically 
equivalent kinetic formulation initialized with 
$N$ originally isolated nodes with a randomly sampled edge added at each discrete time step~\cite{BenNaimPRE05}. Let $T$ denote the number of steps. 
The process is 
parameterized by the 
relative number of introduced edges 
$t=T/N$,   and
typically analyzed in the thermodynamic limit of $N \rightarrow \infty$. 
}
For $t<t_c$ the resulting graph is disjoint, consisting of small isolated clusters (\ie, components) of connected nodes. 
(See Fig.~\ref{figBasics}(c) for an illustration of distinct components.) 
Let $C$ denote the largest component and $|C|$ its size. 
For the ER model the order parameter
$|C|$ undergoes a second order transition at $t_c=1/2$ where, below $t_c$, $|C|$ is of order logarithmic in $N$ and, above $t_c$,
there is a unique largest component with size linear in $N$~\cite{bolloRGbook}.

\section*{The impact of choice} 
At a Fields Institute workshop in 2000, Dimitris Achlioptas introduced 
an extension to the standard process, 
designed to enhance or delay the percolation transition, and which exemplifies the ``power of two choices" as used in 
randomized algorithms~\cite{AzarBroder94,AzarBroder99,Adler98,mitzenmach01}. 
Starting with $N$ isolated nodes, rather than choosing one edge in each discrete time step, choose two candidate edges, denoted $\{e_1, e_2\}$, and examine the consequence of adding each one individually to the graph.  The edge which best satisfies a preset selection criteria is added to the graph and the second candidate edge is discarded for this time step. Selection criteria can include keeping components small
(delaying percolation), or growing a large component as quickly as possible (enhancing percolation).  
The process can also be generalized to consider $m\ge 2$ candidate edges at each time where $m$ is kept constant.  
Such an ``$m$-edge" competitive graph evolution algorithm has come to be known as an {\it Achlioptas Process}.  

Achlioptas Processes were first analyzed by Bohman and Frieze in 2001~\cite{BohmanFrieze} in the context of 
``bounded-size" rules where all components 
of size $K$ or greater are treated equivalently.  For the Bohman and Frieze (BF) process,  
$e_1$ is accepted if it joins two isolated nodes (and $e_2$ rejected), otherwise $e_2$ is accepted (and $e_1$ rejected). Thus, only components of size one (isolated nodes) are distinguished, 
and all components of size $K\ge 2$ 
are treated equivalently. 
A rigorous proof shows that BF delays the percolation transition when compared to ER, 
but the nature of the transition was not investigated~\cite{BohmanFrieze}. 
BF can be modeled as a cluster aggregation process 
based on the Smoluchowski coagulation equation~\cite{smoluchowski,Aldous99}.
This assumes that at each discrete time step, two independent components are merged,
and thus implies that the maximum number of edges possible is $N-1$. 
The error introduced from the violation of this assumption nearing the critical point
can be rigorously analyzed and, as a result, 
it is conjectured that all bounded size rules lead to a continuous phase transition~\cite{SpencWorm}.
Cluster aggregation analysis is a technique 
used in many studies 
discussed throughout this review. 
Note that cluster aggregation processes necessarily end at $t=(N-1)/N$, whereas on an undirected network the maximum edge density attainable is $t=(N-1)/2$. 

\section*{Novel critical properties}
\label{APEP}
Analysis of unbounded size rules is more elusive. The first significant study appeared in 2009~\cite{EPScience} and focused on the Product Rule (PR), an Achlioptas Process defined as follows.  Starting from $N$ isolated nodes, two candidate edges $\{e_1, e_2\}$ are chosen uniformly at random at each discrete time step.  
For $t < t_c$, the largest components are of order logarithmic in $N$ 
and thus, with high probability, the two edges involve four distinct components
with sizes respectively denoted $|C_a|,|C_b|,|C_c|,|C_d|$.  
Let $e_1$ denote the edge which joins the first two components, and $e_2$ the second two.  If  $|C_a|\cdot |C_b| < |C_c|\cdot |C_d|$, then $e_1$ is added to the graph. Otherwise, $e_2$ is added. In other words, we retain the edge that minimizes the product of the two components that would be joined by that edge (see Fig.~\ref{figBasics}(a)).  

A typically realization of a PR process is shown in Fig.~\ref{figBasics}(b),
together with a realization of a 
\ER~(ER)~\cite{ER1959,ER} and a Bohman-Frieze (BF)~\cite{BohmanFrieze} process on a system of size $N=10^6$.
Note that the onset of large-scale connectivity 
is considerably delayed for PR 
and that 
it emerges drastically, going from sublinear to a level approximately equal to the corresponding ER and BF processes during an 
almost imperceptible change in edge density. 
Note, our numerical simulations make use of the ``Newman-Ziff" algorithm for efficient computation of percolation~\cite{NewmanZiff01}.

The 2009 study focused on direct simulation of the PR process~\cite{EPScience}. To quantify the abruptness of the transition, the scaling window as a function of system size $N$, denoted $\Delta_N(\gamma,A)$, was analyzed. This measures the number of edges required for $|C|$ to transition from being smaller than $N^\gamma$ to being larger than $AN$ with typical choices of parameters being $\gamma=A=1/2$.
Systems up to size $N\sim 6\times 10^7$ were studied and the results indicated a sublinear scaling window, $\Delta_N(0.5,0.5) \propto N^{2/3}$, and $t_c \approx 0.888$~\cite{EPScience}. 
The associated change in edge density, $\Delta_N(0.5,0.5)/N \propto N^{-1/3} \rightarrow 0$ as $N\rightarrow \infty$, providing strong,
 yet ultimately misleading, evidence that large-scale connectivity emerges in a discontinuous phase transition. 

Many additional studies followed soon after Ref.~\cite{EPScience}. 
These include PR 
on a lattice~\cite{ZiffPRL09} and on networks with power law degree distributions~\cite{ChoPRL09,RadFortPRL09}.  These studies provided similar evidence for a discontinuous percolation transition. Yet, they also highlighted the existence of scaling behaviors characteristic of second order phase transitions~\cite{ZiffPart2,RadFortPart2}. Many other Achlioptas Processes have now been analyzed, such as rules using the sum rather than product and rules with $m>2$ choices
~\cite{powderkeg09,RDMMclustagg,PRL.107.275703,riordanPRE2012}. 
Similar results of sublinear scaling windows and critical scaling behaviors are observed; see Ref.~\cite{bastas2014review} for a review of many of these processes. 
(Note that models exhibiting a discontinuous jump in an order parameter, but diverging length-scales characteristic of second order transitions,  
are well established for models of ``jamming percolation" on low-dimensional lattices~\cite{schwarz2006onset,PhysRevLett.96.035702,jengPRE2010,cao2012correlated}. These models incorporate spatial correlations intended to capture glassy dynamics in materials.)

Rather than the scaling window, the impact of a {\it single} edge~\cite{naglerNP} provides a more crisp analysis.  Soon after the early studies appeared, it was shown that,
for PR 
and similar $m$-edge processes,
the maximum change in the relative size of the largest component from the addition of a single edge 
decays as a power law with system size,  
$\Delta C_{\rm max} \sim N^{-\beta}$~\cite{naglerNP,Manna}.
Thus the process is continuous as $N \rightarrow \infty$. 
The rate of decay is typically quite small ($\beta=0.065$ for PR~\cite{Manna})
leading to large discrete jumps in systems that are orders of magnitude larger than real-world networks. More details are included later in this review with respect to applications of EP.

Mounting numerical evidence and heuristic arguments indicated that Achlioptas Processes lead, in fact, to 
a continuous phase transition~\cite{daCosta2010,grassberger2011,lee2011continuity,tian2012nature}, 
but with a universality class distinct from any previously observed~\cite{grassberger2011,tian2012nature}.  
(See Ref.~\cite{bastas2014review} for a review of the critical exponents found.)
Finally, in 2011 a rigorous proof by Riordan and Warnke showed that any Achlioptas Process 
leads to a continuous percolation transition~\cite{RWscience2011}.  They proved, in essence, that the number of subcritical components that join together to form the emergent 
macroscopic-sized component 
is not sub-extensive in system size. 
In the words of Friedman and Landsberg, Achlioptas Processes do not lead to the build-up of a 
``powder keg"~\cite{powderkeg09,PRE.83.032101}, 
which is 
a collection of components that contain $cN$ nodes in total where the sizes of the components diverge to $\infty$ as $N \rightarrow \infty$ for some constant $c$. 
Merging the components of such a powder keg would lead to a discontinuous percolation transition.

Yet, Riordan and Warnke showed that, for a random graph, if the number of {\em random} choices $m$ is allowed to increase in any way with system size $N$,  so that $m\rightarrow \infty$ as $N\rightarrow \infty$
(for instance, $m \sim \log(\log N))$, 
then this is sufficient to allow for a discontinuous transition.  For rules not based on randomly chosen $m$ node pairs, however, a discontinuous transition is not guaranteed to occur.
Many EP processes with alternate mechanisms that lead to genuinely discontinuous percolation transitions have now been discovered, as will be discussed later in this review.

\section*{Novel supercritical properties: stochastic staircases}

\begin{figure}[tb]
  \begin{center}
    \includegraphics[width=\textwidth]{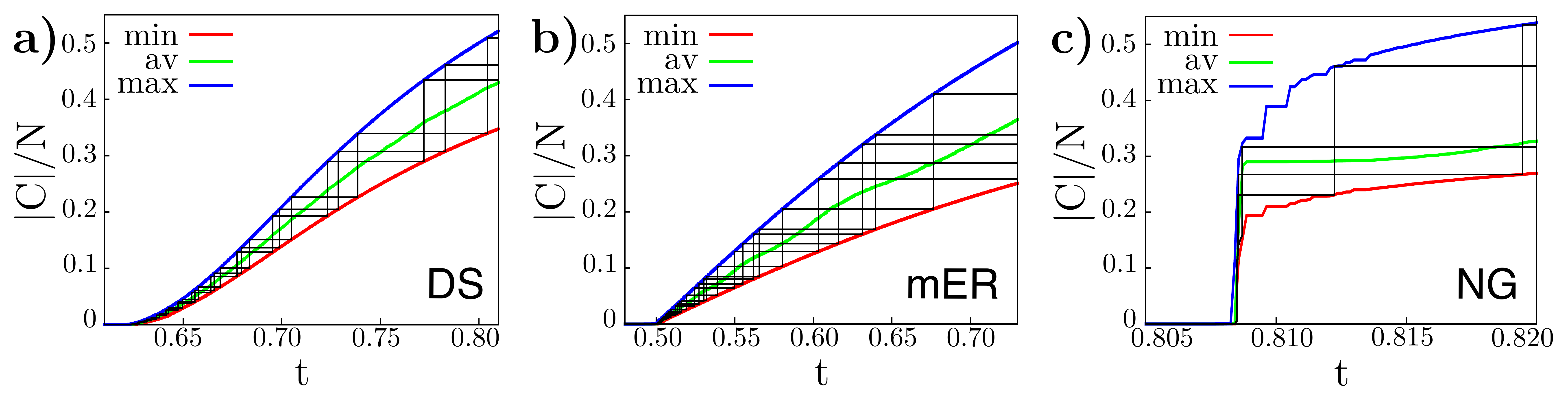}
  \end{center}
  \vspace{-0.2in}
  \caption{  {\bf Explosive Percolation with stochastic staircases.}  
 Models of EP 
can be non-convergent and not self-averaging \cite{riordanPRE2012}.
Shown are realizations of  ``Devil's staircases" of genuinely discontinuous jumps with the relative size 
of the largest component $|C|/N$ as a function of 
 $t$ for several distinct realizations (black lines), together with ensemble average (green line), minimum (red line) and maximum (blue line). 
(a) Devil's Staircase model (DS) analyzed in Refs.\ \cite{naglerPRX} and \cite{schroder2013crackling}.
(b) Modified ER model (mER) from Ref. \cite{riordanPRE2012}.
(c) Nagler-Gutch model (NG) analyzed in Ref. \cite{riordanPRE2012}.
The averages were obtained from 1500 realizations for systems of size $N=2^{30}$.
}
\label{fig:staircases}
\end{figure}

\begin{figure}[tb]
  \begin{center}
    \includegraphics[width=\textwidth]{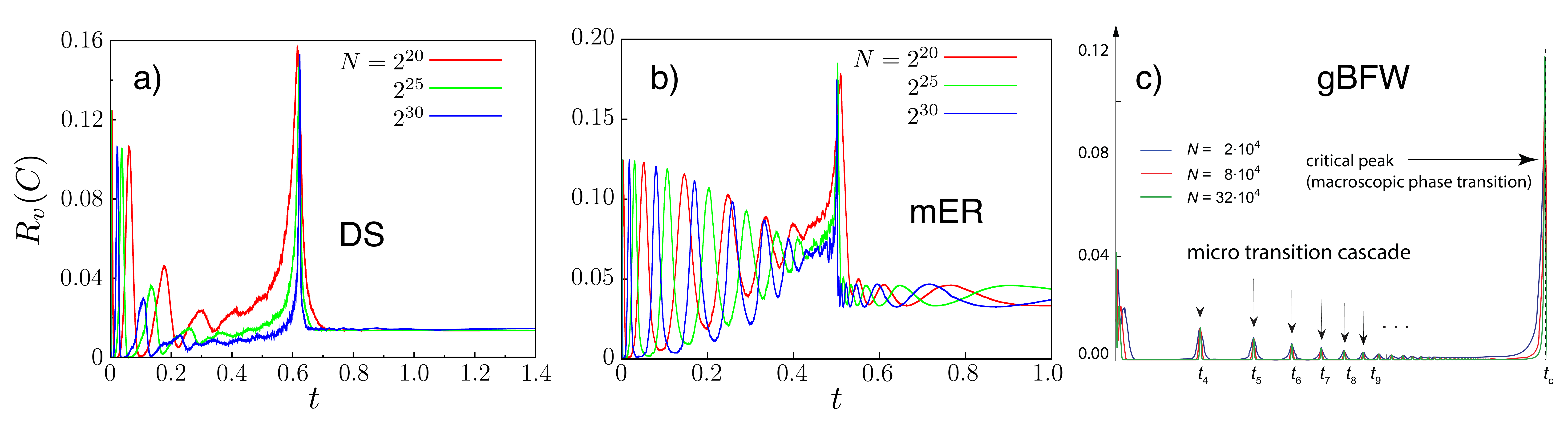}
  \end{center}
  \vspace{-0.4in}
  \caption{{\bf Non-self-averaging in Explosive Percolation.}
For (a) the DS model 
and (b) the modified \ER\ model shown is 
the relative variance $\Rv$ of the largest component in dependence on the link density.
In the supercritical regime the system is 
non-self-averaging characterized by extended regions of $\Rv \neq 0$, for $N\rightarrow\infty$.
For mER, quite remarkably,
 $\Rv$ as a function of $t$ follows intricate patterns such as oscillations
whose amplitude seem to survive in the thermodynamic limit, 
both in the subcritical and supercritical regime.
(c) 
The generalized BFW model (gBFW) studied in Ref.\ \cite{ChenPRL2014} exhibits 
peaks in $\Rv$ at well defined intervals that display a discrete scale invariance and survive in the thermodynamic limit, and moreover predict 
the percolation point.
}
\label{fig:nSA}
\end{figure}

{
Achlioptas Processes (i.e., $m$-edge rules) are continuous. Although a finite realization may show large discrete jumps,
in the limit $N\rightarrow \infty$ the evolution is converges to a smooth, continuous function as illustrated in Fig.~\ref{fig:table}(a). 
But, remarkably, the more general class of ``$k$-vertex rules"  (which consider a fixed number of candidate {\it vertices} rather than {\it edges}) allows for new possibilities. 

To understand the distinction, first consider an $m$-edge rule.
The $m$ vertex {\em pairs} 
are chosen uniformly at random. Hence, (as long as there are at least two components in the system) there is a non-zero probability that all 
candidate edges chosen at a given step have exactly one end-point (i.e.\ vertex) in the 
largest component $C$. Thus, independent of the rule, the probability, $P_{\text{gr}}$,
that the largest component merges with another smaller component is necessarily
non-zero (and even increases during the process as $C$ grows).
This 
results in growth
of the largest component being dominant,
preventing the build up of a powder keg, and leading to
continuous growth of $|C|/N$ in the thermodynamic limit \cite{naglerNP,riordan2012achlioptas}. 

A different mechanism underlies $k$-vertex rules. 
The Devil's staircase rule (DS)  is a 3-vertex rule that preferentially merges components of equal (or similar) size 
or adds an intra-cluster edge~\cite{naglerPRX,schroder2013crackling}.
Hence, regardless of how many of the chosen vertices reside in $C$,
it is {\it impossible} that $C$ merges with a smaller component, meaning $P_{\text{gr}}=0$.
Instead, smaller components merge together 
sometimes overtaking to become  
the new largest component (which can then no longer grow directly).
This condition necessarily implies one or more discontinuous transitions during the process \cite{naglerNP}.
In particular, the DS rule
exhibits a continuous percolation transition at $t_c$, yet exhibits infinitely many discontinuous jumps at $t>t_c$, with the 
``first" such jump within arbitrary vicinity of the initial percolation transition.
Thus continuity at the first connectivity transition and discontinuity of the percolation process can be compatible.  
Moreover, 
the Devil's staircase 
(a Cantor function with discrete jumps) is random, even in the thermodynamic limit, meaning that 
the locations of the jumps are 
stochastic variables, as illustrated in 
Fig~\ref{fig:staircases}(a).

\begin{figure}[h]
  \begin{center}
   \includegraphics[width=0.95\textwidth]{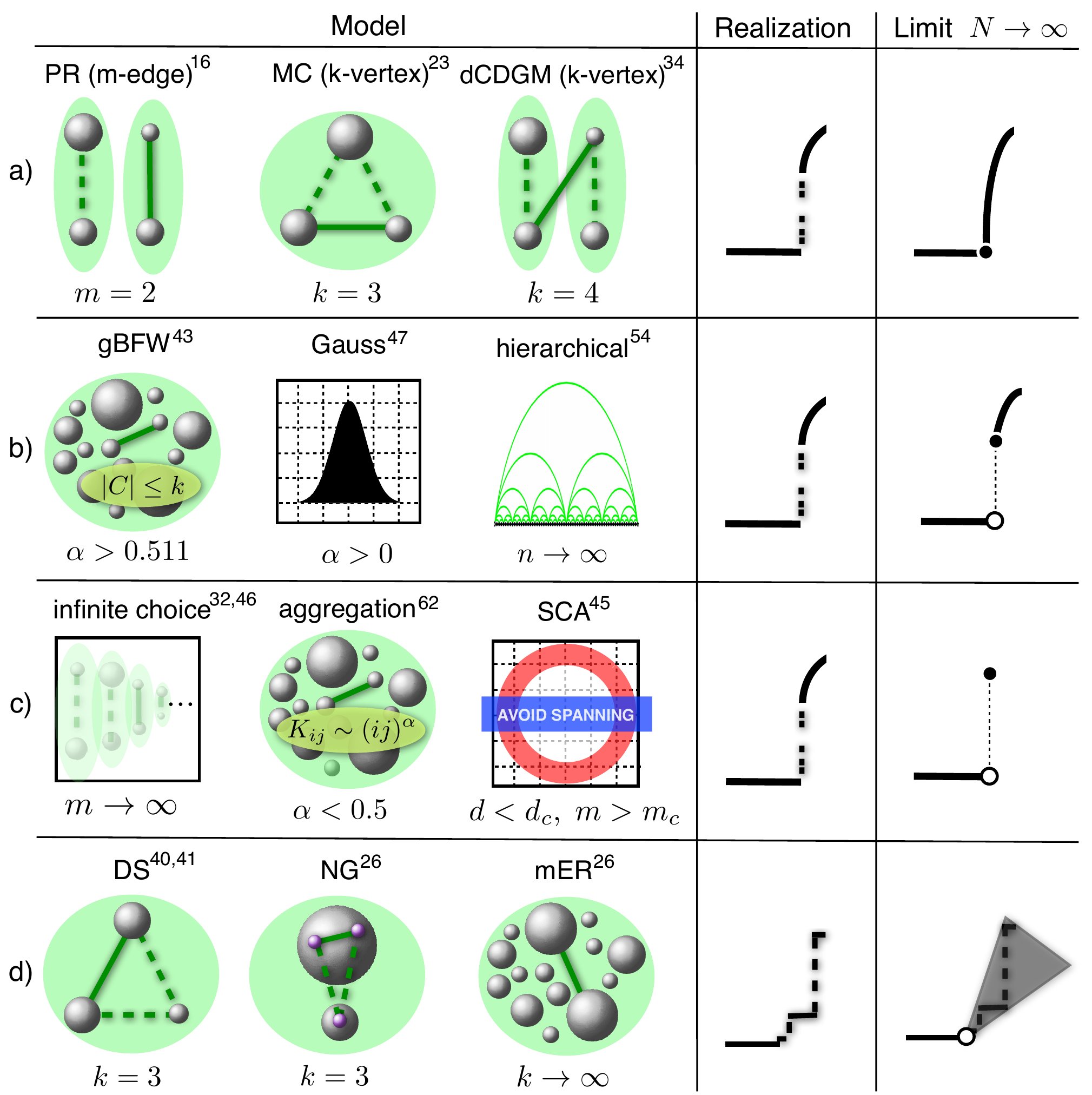}
  \end{center}
  \vspace{-0.2in}
  \caption{
  {\bf Classes of Explosive Percolation.}
  (a) 
  Product Rule, Minimal Cluster Rule and a model named after its creators (dCDGM) are examples of
  Explosive Percolation processes that are continuous in the thermodynamic limit but nevertheless
  exhibit substantial jumps in the order parameter for any finite system.  
  For $m$-edge rules, $m$ links compete for addition. For $k$-vertex rules 
 all possible $k(k-1)/2$ 
 node pairs compete. 
  (b)
  Models that exhibit a single genuine 
  jump in the order parameter $|C|/N$ 
  well in advance of the end of the process. 
  The hierarchical model results from the construction of $n$ generations of long-range bonds, in the limit of $n\rightarrow\infty$.
  (c)
  Models that exhibit a single 
  discontinuous jump in the order parameter $|C|/N$
  at the end of the process  
  resulting in a ``global" jump encompassing the full system. 
  All rules in (a)-(c) delay the onset of percolation and avoid mergers of large clusters.
  (d)
  Non-convergent, non-self-averaging models that exhibit a staircase with genuinely discontinuous steps;
  even in the thermodynamic limit the staircases are stochastic (both the size of the steps and their location).
  For those models, mergers of large components are not explicitly suppressed.
}
\label{fig:table}
\end{figure}
\afterpage{\clearpage}

Other rules where the order parameter $|C|/N$ is ``blurred" in the supercritical regime and does not converge to a function of $t$  in the thermodynamic limit
were reported 
in Ref.~\cite{riordanPRE2012} which preceded Ref.~\cite{schroder2013crackling}.
These include 
a model due to Nagler and Gutch (NG) and a modified \ER\ model (mER)~\cite{riordanPRE2012}.
Both models exhibit 
tremendous variation from one realization to another in the supercritical regime~\cite{riordanPRE2012}, see Fig.~\ref{fig:staircases}(b)-(c).
This behavior is called non-self-averaging
and is quantified by the relative variance of the order parameter $R_v(C)$ over an ensemble of realizations.
For continuous phase transitions it is well known that large fluctuations in $R_v(C)$ 
are observed only in the critical window and that they collapse to a singular peak at $t_c$ in the thermodynamic limit.
Fig.~\ref{fig:nSA} shows the lack of self-averaging
for DS and mER as characterized by elevated values of  
$R_v(C)$ in the supercritical regime. 
Most remarkably, 
large fluctuations in 
$R_v(C)$ can be observed even in the early, {\it subcritical} evolution 
as shown in Fig.~\ref{fig:nSA} for mER and gBFW. These fluctuations can have predictive power as discussed later.

We have illustrated this class of EP phenomena with genuine stochastic staircases 
in Fig.~\ref{fig:table}(d). 
}

\begin{figure}[tb]
  \begin{center}
 \includegraphics[width=\textwidth]{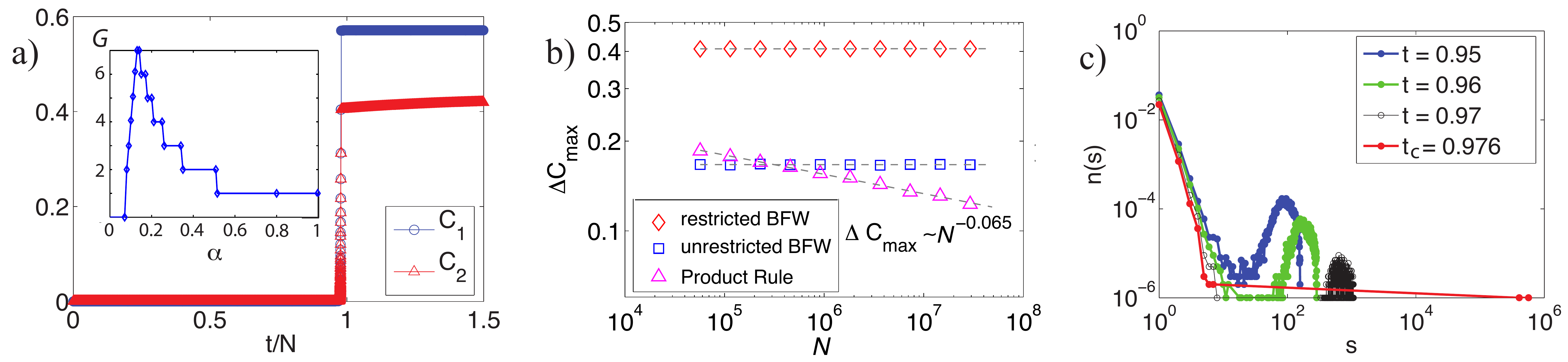}  
  \end{center}
  \vspace{-0.2in}
  \caption{ {\bf Multiple giant components and the powder keg.}
  (a) Multiple giant components, $C_1$ and $C_2$, arise simultaneously for the BFW process. Inset shows the number of stable giant components, $G$, as 
 as a function of $\alpha$
 the asymptotic fraction of edges that must be accepted~\cite{ChenPRL2011}. 
  (b) The maximum impact from a single edge $\Delta C_{\rm max}$ is invariant with system size $N$ for the BFW process, but decays as $N^{-0.065}$ under the Product Rule. (Regardless of whether or not we restrict the BFW process to 
  merge only previously distinct components.) 
   (c) Evolution of the component size distribution, $n(s)$, under the BFW process with $\alpha=1/2$, showing the buildup of the ``powder keg" which merges to become two coexisting giant components at $t_c$.}
\label{figBFW}
\end{figure}

\section*{Lattice models and global percolation phenomena}

Unlike on a random network, an $m$-edge Achlioptas Process on a lattice can yield a discontinuous percolation transition at $t_c$.
Percolation on a lattice is often measured by the emergence of a spanning cluster---a path of activated links that connect sites from one side of the lattice to another.
In Ref.~\cite{cho2013avoiding}, Cho {\it et al.} study what we call the Spanning Cluster Avoidance (SCA) model and show that the emergence of a spanning cluster under
the SCA model
is
discontinuous for a lattice with dimension $d < d_c = 6$ as long as $m \ge m_c = d/(d-d_{\rm BB})$, where $d_{\rm BB}$ is the fractal dimension 
of the ``backbone" (which they calculate analytically and measure numerically).
When $d=2$, $m_c \approx 2.554$ so setting $m=3$ is sufficient for a discontinuous transition.
Yet, an interesting distinction occurs for $m=m_c$ versus $m>m_c$. For $m=m_c$, the discontinuous percolation transition occurs at some intermediate $t_c$ during the process, as shown in Fig.~\ref{fig:table}(b) for other models displaying this class of EP phenomena. In contrast, for $m>m_c$ the process acts globally, 
so when the spanning cluster emerges, it encompasses the entire system. 
Such ``global"  percolation also happens for an $m$-edge Achlioptas Process on a random graph in the limit $m\rightarrow \infty$.
Instead of metric or geometrical confinements, the rule has unrestricted access to the entire collection of components.
There, a giant percolating component only emerges in the final step of the process when only one component remains, as first discussed in~\cite{RozenEPJB2010}. This class of EP phenomena with global jumps is illustrated in Fig.~\ref{fig:table}(c).

\section*{Underlying mechanisms}
\label{disc_models}

\ER~percolation can be considered a form of ``multiplicative coalescence"~\cite{Aldous99}.  From the kinetic perspective, at each discrete time step two vertices are chosen uniformly at random and linked by an edge.
The probability that a randomly chosen vertex is in a particular component of size $j$ is $j/N$. Thus, to first order, the probability that a randomly selected edge merges a particular component of size $j$ with a particular one of size $i$ is proportional to $i j / N^2$. 
(See Ref.~\cite{Aldous99} for more rigorous details.)
As with gravitational attraction, the force between two bodies (\ie, components) is proportional to the product of their masses. 
It suffices to say that, under \ER~evolution, the largest components quickly merge together to form one larger component, hence amplifying the likelihood of that component being included in subsequent edges. 
Such arguments provide the intuition for why there is only one unique giant component in the supercritical regime for \ER~percolation.
  
The appearance of Ref.~\cite{EPScience} led to increased activity in the field and to the discovery of several random graph percolation models that exhibit truly discontinuous transitions.  These models break gravitational coalescence 
allowing instead for a multitude of components with sizes similar to that of the largest component. This creates the necessary ``powder keg"~\cite{powderkeg09} in the sub-critical regime, and can allow for multiple, coexisting giant components in the supercritical regime~\cite{ChenPRL2011}.  

The class of EP phenomena with a discontinuous, but non-global, jump is illustrated in Fig.~\ref{fig:table}(b).
Two of the models shown~\cite{ChenPRL2011,Araujo} work by suppressing the growth of the largest component.
In Ref.~\cite{Araujo} a regular lattice is the underlying substrate and a single edge is examined at a time (\ie, $m=1$).  If a randomly chosen edge would not increase the current size of the largest component then it is accepted. Otherwise it is rejected with a probability function that decays as a Gaussian distribution centered on the average cluster size. Thus, components that are similar in size to the average are favored.  
Clear signatures of a first-order transition are observed, such as bimodal peaks for the cluster size distribution indicating the coexistence of percolative and non-percolative states in finite systems at $t_c$. 
In contrast, the random graph version of this Gaussian model~\cite{KahngArxiv2014} exhibits a discontinuous transition at the end of the process (as illustrated in Fig.~\ref{fig:table}(c)).

In Ref.~\cite{ChenPRL2011} a model previously introduced by Bohman, Frieze and Wormald (BFW)~\cite{BohFriezWormRSA2004} is analyzed.  The model considers a single edge at a time. The edge is added to the graph if the resulting 
component would be smaller than some specified size $k$.  Otherwise, the edge is rejected provided that a stringent lower bound on edge-density is always satisfied. If the edge cannot be rejected, then the cap $k$ is increased incrementally while the lower bound correspondingly decreases as a function of $k$ until reaching an asymptotic limiting value, $\alpha$.  In the original model $\alpha=1/2$ (\ie, asymptotically one-half of all edges must be accepted)~\cite{BohFriezWormRSA2004}.  
Ref.~\cite{ChenPRL2011} shows that this process leads to a truly discontinuous transition in which multiple giant components emerge simultaneously, as shown in Fig.~\ref{figBFW}(a).  Once in the supercritical regime, any edge leading to an increase in the cap size $k$ can be simply rejected and thus the multiple giant components coexist without merging. 

One can generalize BFW (gBFW) by allowing $\alpha$ to vary, 
providing a parameter for tuning the number of 
giant components that emerge at $t_c$ (see inset to Fig.~\ref{figBFW}(a)). 
The critical behavior for $\alpha>0.511$ when only one giant emerges is illustrated in Fig.~\ref{fig:table}(b).
That the maximum 
change in relative size from the addition of a single edge $\Delta C_{\rm max}$ is invariant with $N$ but decays as $N^{-0.065}$ for the Product Rule is shown in Fig.~\ref{figBFW}(b).  
The evolution of the component size distribution as $t$ 
increases is shown in Fig.~\ref{figBFW}(c), illustrating the buildup of the ``powder keg".  Here $n(s)$ denotes the number of components of size $s$ divided by $N$. 
See Ref.~\cite{ChenPRL2011} for details and the explicit BFW algorithm. See Refs.~\cite{ChenPRE2013a,ChenPRE2013b} for details on the supercritical behaviors including stable and unstable coexistence of components, the latter leading to additional discontinuous jumps.

The BFW process 
gives rise to the simple underlying mechanism of growth by overtaking~\cite{ChenEPL2012}.  The growth of the largest component is severely limited
as it can only merge with isolated nodes. 
Instead all significant changes in $|C|$ result from 
two smaller components merging together and overtaking the previous largest component to become the new largest component~\cite{ChenEPL2012}.  

Models that have been shown to lead to a single
discontinuous percolation transition  on random graphs include a restricted \ER~process, 
where one end-point of the edge is chosen uniformly at random and the other is chosen randomly from a restricted set~\cite{panagiotou2011explosive}.  
Ordinary percolation on a hierarchical network can also yield a discontinuous percolation transition
at some intermediate $t_c$ during the process~\cite{boettcher2012}, see Fig.~\ref{fig:table}(b).
Furthermore, 
there is a Hamiltonian formulation that connects evolution via Achlioptas Processes with an equilibrium statistical mechanics process~\cite{Moreira}, 
highlighting 
the role of non-local information in discontinuous percolation.
It was also shown recently that modeling cascading failure on interacting networks via percolation typically involves a discontinuous transition 
from global connectedness to disintegration of the network~\cite{buldyrev2010}.

Although percolation considers the evolution of the network structure, 
EP has also motivated exploration of 
dynamical processes taking place on a fixed network structure, such as the ``explosive Ising" model~\cite{angst2012explosiveIsing} and ``explosive synchronization"~\cite{explosiveSyncPRL2011}.
The latter was first shown in a network of oscillators when the natural frequency of each oscillator is positively correlated with its degree.  A recent study revealed that suppressing the formation of large clusters is the common mechanism underlying
all now explored models of explosive synchronization~\cite{explosiveSyncPRL2015},
linking the mechanism with EP. 

Notably, there have long been 
models of percolation known to show discontinuous transitions, such as $k$-core percolation and models of jamming on low-dimensional lattices~\cite{schwarz2006onset,PhysRevLett.96.035702,jengPRE2010,cao2012correlated}.  Mechanisms underlying these processes are primarily cooperative interactions~\cite{bizhani2012}
and correlated percolation~\cite{cao2012correlated}. 
Refs.~\cite{cao2012correlated,bizhani2012} include interesting discussions connecting these known models to the more recent work on EP, highlighting lattice models, generalized epidemic models and the statistical mechanics of exponential random graphs.

\section*{Explosive Percolation in real-world networks}
\label{realworld}

\subsection*{Explosive percolation in finite systems}
The rigorous proof by Riordan and Warnke~\cite{RWscience2011} shows that in the limit $N\rightarrow \infty$ the scaling window is linear in system size $N$, but numerical evidence on systems up to size $N \sim 10^7$ indicates 
the window is sublinear~\cite{EPScience}.  Thus, there must be a crossover length, $N^*$, where the system becomes large enough that actual realizations show convergence to the asymptotic limiting behavior.  
A method for estimating the crossover length is to model the expected evolution of a network using cluster aggregation equations, such as the Smoluchowski coagulation equation~\cite{smoluchowski,Aldous99} discussed earlier, which is a ``mean-field" analysis
over the ensemble of all possible random graphs.  See Ref.~\cite{Aldous99} for details of the
convergence and concentration assumptions underlying this approach.  
Cluster aggregation approaches to general percolation provide useful analytical tools~\cite{BenNaimPRE05,krapivsky2010book}, which have been quite conducive for modeling EP processes~\cite{ChoClusterAgg,RDMMclustagg,Manna}.  

In their arguments establishing the continuity of percolation under Achlioptas Processes, da Costa {\it et al.} analyze cluster aggregation models related to the Product Rule and show that the largest component obeys a scaling relation $|C|/N \sim (t-t_c)^{0.0555}$ for $t$ just above $t_c$~\cite{daCosta2010}.  
This indicates unusually rapid, albeit continuous, growth. 
As discussed earlier, the maximum impact from the addition of a single edge for such processes obeys the relation $\Delta C_{\text{max}} \sim N^{-\beta}$, which for very small values of $\beta$ coincides with the scaling of the largest component
$|C|/N \sim (\Delta t)^\beta =  (1/N)^\beta$, as the addition of a single edge corresponds to $\Delta t = 1/N$. 
For $t < t_c$, by definition $|C|/N \rightarrow 0$. As we pass into the critical regime $|C|/N \sim 
(1/N)^\beta$. 
This means that, for a system of size $N=10^{1/\beta}$, the addition of a single edge causes the order parameter to exhibit a discrete jump equal to ten percent of the system size, $\Delta |C|/N = 0.1$.  For a process with $\beta=0.0555 \approx 1/18$, the crossover length $N^* > 10^{18}$.  The thermodynamic limit is extremely relevant when considering phase transitions in physical materials, where system sizes are on the order of Avogadro's number, $N\sim 10^{23}$. But real-world networks, such as the internet, the world-wide airline network, online social networks, gene interaction networks, etc., are all considerably smaller than $10^{18}$.  
Although fixed choice Achlioptas Processes yield continuous transitions in the thermodynamic limit, such processes yield significant discrete jumps in the realm of real-world networks.

\subsection*{Modular networks}

Several studies show that the paradigm of EP can be useful for understanding the evolution of modular networks and community structure. 
In Ref.~\cite{RozenEPJB2010} Rozenfeld {\it et al.} consider an evolutionary process on the Human Protein Homology Network. The general belief is that proteins evolve via duplication-mutation events from ancestral proteins, and it has been shown that more similar (\ie, homologous) proteins organize into network modules~\cite{MediniPLOS2006}.  Using data on the human protein homology network, Rozenfeld {\it et al.} consider an evolutionary process initialized with all the proteins disconnected and with edges between the most similar proteins added sequentially.  This leads to the emergence of many large isolated components of tightly connected nodes, which eventually link together with the addition of just a few intercomponent edges so that global connectivity emerges in an explosive manner.  
As the authors remark, the emergent structure is similar to the dense connectivity within a community and the weak links between communities suggested by Grannoveter for social systems~\cite{granovetter1973}.

Pan {\it et al.} show in Ref.~\cite{PanPRE2011} that monitoring the evolution of an EP process on a network reveals  information about the underlying structure. They consider empirical data from two real-world social networks: one is a mobile phone call network, the other is a co-authorship networks of scientists.  Initially, all the empirical edges are considered ``unoccupied" and an Achlioptas Process is used to sequentially ``occupy" edges.  They show that, at $t_c$, the component structure reflects the underlying community structure of the network.  Thus, applying such graph evolution processes to data from real-world networks can provide a potential tool for uncovering unknown, underlying structures. 

In Ref.~\cite{bounovaThesis}, Bounova studies the first-year growth of many distinct language wikipedias (e.g., French, Italian, Chinese, Esperanto).  Each wikipedia is a network of articles connected via hyperlink edges. Most of the languages exhibit the same general pattern of evolving a collection of large disconnected components, with each component focused on a distinct topic. Similar to the ``powder keg", these distinct components 
quickly link together over the course of a few days, leading to large discrete jumps in the size of the largest component.

\subsection*{Disordered media}

Standard formulations of percolation have been used to model many properties of materials and disordered media, 
such as electrical and thermal conductivity, flow through porous media, and polymerization.  
Explosive Percolation offers a novel ingredient, namely suppressing the growth of the largest components and instead creating many components of uniform size. This allows us to extend percolation models to systems that have not been previously amenable to such treatment.  

For example, consider the seminal model of diffusion-limited cluster aggregation~\cite{witten1981}.  Here clusters move via Brownian motion so that the velocity of a cluster is inversely proportional to the square root of its size and thus larger clusters move considerably more slowly.  In Ref.~\cite{ChoKhangPRE2011}, Cho and Kahng show that diffusion-limited cluster aggregation can be mapped onto the framework of EP.  They consider clusters moving on an underlying two-dimensional lattice via Brownian motion, and whenever two clusters become nearest neighbors, they merge into one larger cluster. 
They show that Brownian motion suppresses of the mobility of the largest clusters, impeding their growth, and leading to the discontinuous emergence of a giant cluster as a function of
the number of aggregation events.
They also consider a generalized Brownian motion where the velocity is inversely proportional to the mass of the cluster to a 
power $\eta$ and map out the tricritical point separating discontinuous from continuous emergence as a function of $\eta$. 

Schr{\"o}der {\it et al.} introduced a generalization of the DS model, called  ``fractional percolation" where
the merging of components with substantially different sizes is systematically suppressed and 
components are prefentially merged whose size ratio is close to a fixed target ratio $f$ \cite{schroder2013crackling}. 
For any target ratio $f$ (no matter how small) they show that this leads to a series of multiple discontinuous jumps in the supercritical regime. %
The sizes and locations of the jumps are randomly distributed, similar to crackling noise observed in materials, such as when a sheet of paper is crumpled.
Their framework links EP with phenomena that exhibit crackling noise, are non-self-averaging, 
and exhibit power-law fluctuations resembling  Barkhausen noise in ferromagnets. 

Recently, the electric breakdown of substrates on which highly conducting particles are adsorbed and desorbed has been identified as a promising candidate of an experimental realization exhibiting a truly discontinuous percolation transition \cite{HerrmannPRL2014}.

The behaviors of nanotubes are often modeled via standard percolation where 
the emergence of percolating paths in bundles of nanotubes captures the transition from insulator to conductor~\cite{explosiveNanotubesPRE2010}. 
But, in Ref.~\cite{explosiveNanotubesPRE2010}, Kim {\it et al.} show that EP processes are more realistic models as observations of real-world systems show that the sizes of the bundles are uniform.  Similar to EP  processes (and unlike regular percolation), the growth of larger bundles is suppressed and the transition becomes extremely abrupt. The transition show hysteresis, as is expected for first order transitions~\cite{explosiveNanotubesPRE2010}.

\section*{Recent developments}

The cluster aggregation approach that informs much of the work reviewed here also allows us to study competitive percolation processes on growing networks.  Note that in all the percolation models discussed thus far, $N$ is fixed and the graph evolves via edge arrival. In a seminal study appearing in 2001, the impact of node arrival on the \ER~process was analyzed~\cite{callaway2001}.  Starting from a few seed nodes, a new node arrives at each discrete time step and, with probability $\delta \le 1$, an edge selected uniformly at random is added to the graph.  This leads to a infinite order percolation transition~\cite{callaway2001}.  Following the same procedure, but using the ``Adjacent Edge" Achlioptas Process~\cite{RDMMclustagg} for edge addition, 
considerably delays the onset of the percolation transition but retains the 
smooth, {\it infinite} order transition~\cite{vikramPRE2013}.  Thus, network growth via node arrival allows for a significantly delayed percolation transition yet can mitigate the abrupt, explosive nature that typically results from delay interventions~\cite{vikramPRE2013,growthEPL2013}.  

Also shown recently is that microscopic patterns in the early evolution of percolation processes can be used to predict the location of the critical point~\cite{ChenPRL2014}. 
In particular, the gBFW process exhibits peaks in relative variance at well-defined values of $t_i$ (with $i$ an integer),
which survive in the thermodynamic limit, see  Fig.~\ref{fig:nSA}(c).
The positions of the peaks $t_i$ obey a 
discrete scale invariance \cite{Sornette1998} 
(meaning that scale invariance only holds for a discrete set of magnification factors). 
We can predict the
critical point $t_i\rightarrow t_c$ 
from the discrete scaling relation~\cite{ChenPRL2014}.
Non-self-averaging behaviors can thus provide a powerful predictive tool.

Very recently a strict scaling theory for a wide class of Achlioptas Processes was developed using the cluster aggregation approach which produces the full set of scaling functions and critical exponents~\cite{daCostaPRE2014}. 
Even more recently, 
the necessary conditions that a cluster merging process must satisfy to produce a discontinuous percolation transition were established, both for transitions of the type shown in Fig~\ref{fig:table}(b) and (c)~\cite{KahngArxiv2014}.  The key ingredient involves whether symmetry is preserved or broken during cluster merging. 
Finally, we note that novel approaches based on analyzing the matrix describing non-backtracking walks on graphs have recently proven to help in determining the position of the percolation point, the size of the percolating cluster, and the average cluster size~\cite{PRL.113.208701,PRL.113.208702,PRL.91.010801}. Such approaches may become helpful for arbitrarily complex percolation models in the future, including for explosive models.

\section*{Future Directions}

\label{future}

There are many directions for future work on the topic of Explosive Percolation, ranging from more theoretical considerations to more practical aspects of
how these processes can help us model, control, and understand real-world systems. One direction is how EP processes can be used for creating and analyzing modular networks, furthering the initial studies~\cite{RozenEPJB2010,PanPRE2011}. Ordinary percolation on hierarchical lattices leads to an EP transition~\cite{boettcher2012} and may also show interesting connections to community structures and clustering phenomena. There is also very limited work concerning EP on directed networks with the work thus far focused on $m$-edge Achlioptas Processes~\cite{SquiresPRE2013}.

A more novel consideration is the range of supercritical properties observed in EP processes, such as multiple giant components and 
stochastic staircases. 
Some mechanisms that yield EP (\eg, growth by overtaking) lead to one phase transition and stable coexisting giant components. Yet, other mechanisms result in unstable coexistence and a family of discontinuous, supercritical transitions. 
Moreover, the 
fact that multiple giant components arise in percolation is surprising~\cite{SpencerAMS2010}, given the gravitational attraction underlying classic processes such as \ER.  
Understanding which mechanisms lead to stable and unstable coexisting giants may provide insight into the evolution of modular networks, such as social networks, and also provide a potential mechanism for controlling gel sizes during polymerization when  multiple disconnected polymer gels can be desirable~\cite{Krapiv2005}.  
Other real-world systems that may benefit from, and contribute to, deeper understanding of EP processes include
diffusion-limited cluster aggregation and properties of nanotubes and nanowires. 
The lack of self-averaging throughout the process and non-convergence in the supercritical regime that is observed for many EP processes challenges our current notions of percolation.  Even for the basic Product Rule process, there remain many open questions as detailed in~\cite{riordan2012achlioptas,riordan2014evolution}.

From a conceptual perspective, the insights gained from EP processes may help us understand how to better manage and control networks. With our increasing reliance on interdependent systems of networks, from electric power grids, to computer networks, to transportation networks and global financial networks, there is increasing need to understand the systemic risk underlying these engineered networks. Often human operators or regulators intervene with a network's functions or structure in an attempt to delay an undesirable outcome, such as a leak in a dam or a crash in a financial market. 
Such delay interventions can sometimes be successful. Yet, at other times can lead to unanticipated and disastrous failures. 
EP processes provide a new paradigm for modeling the consequences of repeated, small interventions intended to delay a catastrophe and EP has been proposed as a paradigm for modeling modern engineered and financial systems~\cite{ETHconference2012,helbing2013globally}. 

The fields of percolation and Explosive Percolation are extremely active.
Many interesting papers have appeared that could not be included in this brief review which is focused on the novel properties and classes of EP phenomena. For recent reviews of advances and challenges in the general field of percolation see Ref.~\cite{AraujoPercReview2014,SaberiPhysReports2015}. See Ref.~\cite{bastas2014review} for a comprehensive review of models displaying EP.

\vspace{0.2in}
\noindent
{\bf Acknowledgements:} 
We thank M. Schr\"oder and A. Witt for valuable discussions and assistance in the preparation of the figures,
and Q. D'Souza and L. Nagler-Deutsch for invaluable input.
R.M.D. gratefully acknowledges support from the U.S. Army Research Office MURI Award No.~W911NF-13-1-0340 and Cooperative Agreement No. W911NF-09-2-0053 and the Defense Threat Reduction Agency Basic Research Award HDTRA1-10-1-0088. 

\bibliographystyle{unsrt}
\bibliography{/Users/raissa/Documents/Bib/prodRule,/Users/raissa/Documents/Bib/netGrowth}{}

\end{document}